\documentclass[aps,reprint,pra,superscriptaddress,floats,floatfix,nobibnotes,longbibliography]{revtex4-2}

\usepackage[utf8]{inputenc}
\usepackage{amsmath}
\usepackage[caption = false]{subfig}
\usepackage{graphicx,epstopdf}
\usepackage{blindtext}
\usepackage[table,xcdraw]{xcolor}
\usepackage{lipsum}
\usepackage{amsfonts}
\usepackage{bbm}
\usepackage{amssymb}
\usepackage{enumerate}
\usepackage{color}
\usepackage{latexsym}
\usepackage{physics}
\usepackage{amstext}
\usepackage{times}%,txfonts}
\usepackage{threeparttable}
\usepackage{graphicx}
\usepackage[colorlinks=true,linkcolor=blue,urlcolor=blue,citecolor=blue]{hyperref}
\usepackage{orcidlink}
\usepackage{soul}

\usepackage[normalem]{ulem}

\usepackage[ragged]{sidecap}

\newcommand{\iu}{\textnormal{i}}

\newcommand{\UFSCar}{Departamento de Física, Universidade Federal de São Carlos (UFSCar), 13565-905 São Carlos, São Paulo, Brazil}

\newcommand{\Unesp}{Universidade Estadual Paulista (UNESP), Instituto de Ciências e Engenharia, 18409-010 Itapeva, São Paulo, Brazil}

\newcommand{\UEPG}{Universidade Estadual de Ponta Grossa (UEPG), Departamento de Física, 84030-900 Ponta Grossa, PR, Brazil}

\newcommand{\Title}{Building Block For Universal Continuous Variables Computation In Superconducting Devices}

\newcommand{\dagg}[1]{{#1}^\dagger}

\begin{document}

\title{\Title}

\author{Bruno A. Veloso \orcidlink{0000-0002-5935-7413}}
\email{brunoavn@df.ufscar.br}
\affiliation{\UFSCar}

\author{Ciro Micheletti~Diniz \orcidlink{0000-0002-7602-0468}}
\affiliation{\UFSCar}

\author{Luiz O. R. Solak\,\orcidlink{0000-0002-4760-3357}}
\affiliation{\UFSCar}
\affiliation{CESQ/ISIS (UMR 7006), Université de Strasbourg and CNRS, 67000 Strasbourg, France}

\author{Antonio S. M. de Castro\,\orcidlink{0000-0002-1521-9342}}
\affiliation{\UEPG}

\author{Daniel Z. Rossatto \orcidlink{0000-0001-9432-1603}}
\affiliation{\Unesp}

\author{Celso J. Villas-Bôas
\orcidlink{0000-0001-5622-786X}
}
\affiliation{\UFSCar}

%\date{\today}

\begin{abstract}
    Continuous variable (CV) quantum computation offers an alternative to
    qubit-based computing by exploiting the infinite-dimensional Hilbert space
    of bosonic modes. Despite recent progress, superconducting platforms have
    yet to demonstrate a scalable architecture capable of universal computation.
    Here, we design and numerically simulate a two-layer superconducting
    architecture that implements all five interactions of the universal CV gate
    set (rotation, displacement, squeezing, Kerr, and beam splitter) within
    experimentally accessible regimes. To this end, we employ a DC-SQUID as the
    bosonic mode, a fluxonium qubit to mediate nonlinear interactions, and two
    ancillary qubits that enable Gaussian and multi-mode operations. By tuning
    fluxes and frequencies, we achieve high fidelities ($\geq
    98\%$) across all gates within state-of-the-art parameter ranges.
    The modular nature of the design allows straightforward scaling,
    establishing a feasible pathway toward high-fidelity, universal CV quantum
    computation based on superconducting circuits.
\end{abstract}

\maketitle

\section{Introduction}
The race toward developing quantum processing units (QPUs) has achieved notable progress in recent years, with several platforms proposed for quantum hardware, including quantum dots~\cite{Loss1998}, trapped ions~\cite{Cirac1995, Haffner2008}, photonic circuits~\cite{OBrien2007} and Majorana
particles~\cite{Aguado2020, Microsoft2025}. Among these platforms, superconducting qubits~\cite{Kjaergaard2020, Arute2019} stand out due to their scalability and controllability, with some superconducting QPUs already performing sophisticated tasks~\cite{Arute2019, Zhu2022, Morvan2024}.

Traditionally, quantum computing is formulated using discrete-variable systems (qubits), which are manipulated through a set of quantum gates that enable universal computation, a paradigm analogous to classical digital computation. Nonetheless, an alternative to this approach is continuous-variable (CV) quantum computation, which encodes information in the infinite-dimensional Hilbert space of a bosonic system~\cite{Lloyd1999}. In this paradigm, universality requires Hamiltonians that are arbitrary-order polynomials in the mode quadratures. Those can be built through five interactions that compose the universal gate set for CV computation~\cite{Lloyd1999, Braunstein2005, Fukui2022}: three Gaussian operations (rotation, displacement, and squeezing), one higher-order non-Gaussian operation (beyond second order) and one interaction ensuring controlled information exchange between adjacent modes (beam splitter).

Among the possible non-Gaussian operations, the Kerr nonlinearity is particularly attractive for superconducting implementations, as it arises naturally from the anharmonicity of transmon qubits~\cite{Krantz2019, Kjaergaard2020}. Taking this as the non-Gaussian gate, the universal gate set reads,
\begin{subequations}\label{eq:CV_set}
\begin{eqnarray}
\mathcal{R}(\theta) &=& \exp[\iu \theta \dagg{a}_{k} a_k],\label{sub_eq:R} \\
\mathcal{D}(\alpha) &=& \exp[(\alpha a_k^\dagger   - \alpha^*
a_k)],\label{sub_eq:D}\\
\mathcal{S}(\xi) &=& \exp[-\frac{1}{2}( \xi \dagg{a}_{k}{}^{2}  - \xi^{*}a_k^{2}
)], \label{sub_eq:S} \\
\mathcal{K}(\chi) &=& \exp[\chi {(\dagg{a}_k a_k )}^2],\label{sub_eq:K}\\
\mathcal{B}(\beta) &=& \exp[-\iu\beta( e^{\iu\phi}\dagg{a}_k a_{k+1} + e^{-\iu
\phi} a_k \dagg{a}_{k+1})] \label{sub_eq:B},
\end{eqnarray}
\end{subequations}
where $\dagg{a}_{k}(a_{k})$ are the creation (annihilation) operators for the $k$-th mode in the QPU, while $\theta$, $\alpha$, $\xi$, $\chi$, $\phi$ and $\beta$ are tunable parameters for implementing the rotation,
displacement, squeezing, Kerr, and beam splitter operations, respectively. To ease notation, from now on, the $k$ indexes will be suppressed for all operations that do not require multimodal interactions.

Universal CV quantum systems have been achieved both in optical devices~\cite{Fukui2022, Anai2024, Braunstein2005, Arrazola2021} and in trapped ions~\cite{OrtizGutirrez2017}. More recently, superconducting implementations have emerged~\cite{Hillmann2020, Eriksson2023, He2023, Iyama2024} that implement certain interactions of the universal gate set. However, current CV devices in superconducting circuits still fail to achieve full universality while meeting fundamental criteria for practical quantum computation~\cite{DiVincenzo1996}.
Specifically, recent works~\cite{Hillmann2020, Eriksson2023} have explored nonlinear resonators to encode CV modes, relying on external fluxes to implement interactions. However, the geometry of superconducting resonators poses an intrinsic limitation to scalability, restricting the number of devices that could be coupled to form a multimodal device~\cite{LevensonFalk2025}.

The present work addresses the aforementioned problems by proposing a scalable superconducting device capable of controllably implementing a universal CV-gate set. By exploiting the tunable parameters of the proposed model, within the range of state-of-the-art superconducting devices, we can control all interactions from the universal CV set with fidelities exceeding 98\% for all gates, reaching up to 99\% for Gaussian operations. Furthermore, by leveraging recent works in multilayer superconducting circuits~\cite{Brecht2016,Kosen2022,Conner2021, ciro_reset_2025}, we employ a three-dimensional architecture that overcomes geometric constraints, enabling scalability beyond what is achievable with planar designs.
\begin{figure*}[ht]
    \centering
    \includegraphics[width=17.2cm]{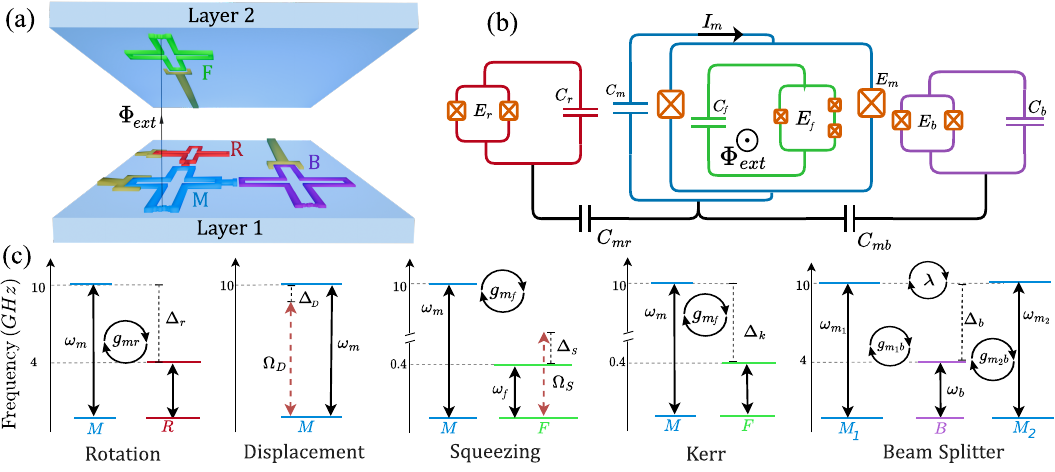}
    \caption{\textbf{(a)} Schematic representation of the building block for universal CV quantum computation. The circuit is composed of a two-layer circuit. Layer 1 comprises a DC-SQUID ($M$) that will encode our continuous
    variables and the auxiliary qubits ($R$ and $B$) responsible for the rotation and beam splitter interactions. On Layer 2, a fluxonium qubit ($F$) is coupled via an external magnetic flux $\Phi_{\textnormal{ext}}$ to the DC-SQUID, allowing the tuning of the second-order interaction between these devices. The arms connected to each device stand for the control lines that apply the fields that control the device frequencies. \textbf{(b)} Circuit diagram of
    schematics in (a). Each superconducting element $i=\{m,f,r,b\}$ is composed of a superconducting ring with two or more Josephson junctions with
    Josephson energies $E_{i}$ and capacitors $C_{i}$. We also have  capacitive coupling between the mode and the auxiliary qubits in Layer 1. We assume no capacitive coupling between $F$ and the auxiliaries $R$ and $B$ due to the physical separation of the elements. The application of each interaction depends on our ability to tune the devices and apply external pulses via the control lines. The energy configuration for enabling each interaction can be seen in \textbf{(c)} where the relevant pulses $\Omega_i$ and detunings $\Delta_i$ are controlled to generate the interactions. The frequency values in each diagram stand for the ones used in our simulations.}
    \label{fig:system}
\end{figure*}

\subsection{Physical Device}

The building block proposed in this work is depicted in
Fig.~\ref{fig:system}(a) and (b). It consists of a two-layer superconducting circuit where each element operates at a tunable frequency $\omega_k$ ($k=\{m,f,r,b\}$). The first layer contains a DC-SQUID (mode $M$) and two auxiliary tunable-frequency qubits: $R$ (rotation qubit) and $B$ (beam splitter qubit). The second layer houses a fluxonium qubit $F$~\cite{Bao2022}. The bosonic mode $M$ is described by annihilation and creation operators $a$ and $a^\dagger$, while the qubits $F$, $R$, and $B$ are treated as two-level systems~\cite{Hu2023} described by Pauli matrices $\sigma^{(i)}_x = \ket{e}_{i}\bra{g}_{i} + \textnormal{H.c.}$ and $\sigma^{(i)}_z=\ket{e}_{i}\bra{e}_{i}-\ket{g}_{i}\bra{g}_{i}$, where $\ket{g}_{i}$ ($\ket{e}_{i}$) denotes the ground (excited) state.

The fluxonium qubit $F$ in the top layer is inductively coupled to the DC-SQUID via an external magnetic flux $\Phi_{\textnormal{ext}}$, following the circuit architecture of Refs.~\cite{Bertet2005, Felicetti2018}. In this scheme, a quadratic coupling controlled by $\Phi_{\textnormal{ext}}$ enables the implementation of the nonlinear operations ($\mathcal{S}$ and $\mathcal {K}$) in our gate set. The corresponding Hamiltonian reads (using  $\hbar=1$)
\begin{align}
\label{eq:Hamiltonian_MF}
 H_{MF} &= \omega_{m}\dagg{a}a+\frac{\omega_{f}}{2}\sigma_{z}^{(f)}+g_{mf}(\dagg{a} +a)^{2}\sigma^{(f)}_{x},
\end{align}
where the inductive coupling strength $g_{mf}$ is tunable via the external flux $\Phi_{\textnormal{ext}}$.

Within the first layer, the two auxiliary qubits $R$ and $B$ are capacitively coupled to the DC-SQUID in a usual Jaynes-Cummings interaction~\cite{Rasmussen2021}. These standard couplings will be used to control the operations $\mathcal{R}$, $\mathcal{D}$, and $\mathcal{B}$. Combining both subcircuits, the total Hamiltonian for a single building 
block is given by \begin{align}\label{eq:Hamiltonian_complete}
H = H_{MF} + \sum_{i=r,b} \left[\frac{\omega_{i}}{2} \sigma_{z}^{(i)} + g_{mi}\left(\dagg{a} + a\right) \sigma_{x}^{(i)}\right],
\end{align}
where $g_{mi}$ are the capacitive coupling strengths.

The two-layer architecture employed in our work is essential,  since the quadratic coupling $g_{mf}$ required for nonlinear operations is activated by threading an external magnetic flux $\Phi_{\textnormal{ext}}$ through both the DC-SQUID $M$ and the fluxonium $F$, following the method demonstrated in Ref.~\cite{Felicetti2018}. While this multilayer approach adds fabrication complexity, 3D superconducting architectures~\cite{Brecht2016, Rosenberg2017} have seen significant advances in recent years, including improved inter-layer coupling techniques~\cite{Kosen2022} and flip-chip integration methods~\cite{Conner2021, Li2021, Luo2025}. These developments position multilayer designs as a promising pathway toward scalable quantum processors~\cite{Kosen2022, Luo2025}, making our proposed architecture experimentally feasible with current technology. 

Given the three-dimensional aspect of the circuit, $F$ will be physically distant from qubits $R$ and $B$; therefore, the coupling between them is assumed negligible~\cite{Rosario2023, Santos2023}. It should be noticed that, apart from the mode, we assume all devices to be ideal two-level systems. This approximation is valid provided that higher excited states remain negligibly populated during the system evolution. For the auxiliary qubits $R$ and $B$, this is ensured by maintaining them in the ground state~\cite{Hu2023}. For the fluxonium qubit $F$, despite being coherently driven during squeezing operations, the large anharmonicity characteristic of this device ($\alpha_f/2\pi \gtrsim 500$\,MHz~\cite{Bao2022}) suppresses transitions to higher levels under the drive amplitudes considered in this work.

Superconducting circuits have already demonstrated high scalability, with circuits having dozens or even hundreds of qubits~\cite{Arute2019,Zhu2022}. Such scalability and the developments in multilayer superconducting circuits~\cite{Brecht2016, Kosen2022, Conner2021} are the fundamental characteristics allowing us to scale the circuit. For instance, a set of four superconducting devices ($M_k, F_k, R_k, B_k$) constitutes the k-th building block of the CV QPU. By coupling another DC-SQUID ($M_{k+1}$) to the auxiliary qubit ($B_{k}$) and replicating the architecture shown in Fig.~\ref{fig:system}(a), the number of controllable modes increases by one. Iterating this procedure yields an $N$-mode QPU.

Once the circuit has been established as described in Fig.~\ref{fig:system}, the unitary evolution of an initial state $\ket{\psi_0}=\ket{\psi}_{m}\otimes\ket{\psi}_{f}\otimes\ket{\psi}_{r}\otimes\ket{\psi}_{b}$ is given by $\ket{\psi(t)} = U(t) \ket{\psi_0}$, where $U(t) = \exp(-\iu Ht)$ is the evolution operator of the system. To construct the universal CV gate set, we must determine the range of parameters of $H$ that ensures $U(t)$ performs each operation in Eq.~(\ref{eq:CV_set}). For this purpose, we will control both the frequencies and the interaction times.

\section{Engineering Interactions}\label{sec:Engineering_Interactions}
To implement each operation, detunings and external drives must be applied to the circuit elements so that the effective dynamics of the mode $M$ matches the operations in Eq.~(\ref{eq:CV_set}). To evaluate the fidelity of each interaction $i$, we define $H_{i}$ as the complete Hamiltonian of the circuit. The system then evolves for an interaction time $\tau$ until it reaches the state $\exp(-\iu H_{i} \tau)\ket{\psi_0}$. We then compare this result with the expected state obtained from the corresponding effective Hamiltonian. To quantify the similarity between the effective and total results, we compute the fidelity defined as~\cite{Uhlmann1976, qutip2024}
\begin{equation}
    \mathcal{F}(\rho_1, \rho_2) =
    \left( \mathrm{Tr} \sqrt{ \sqrt{\rho_1} \, \rho_2 \, \sqrt{\rho_1} } \right)^2,
\end{equation}
where $\rho_{1}$ and $\rho_2$ represent the target and obtained final states after tracing out the auxiliary circuit elements, leaving only the mode.

For each operation, only the relevant elements should interact with the mode $M$ at any given time. To prevent residual interactions from affecting fidelities, we tune non-essential elements to frequencies far from resonance with $M$. By maintaining off-resonant frequencies and the qubits in their ground states~\cite{Santos2023, Hu2023}, we can assume that only the relevant elements are interacting at a given time. In particular, $g_{mf}$ can be turned off by tuning the external flux~\cite{Felicetti2018, supp_mat}.

To evaluate the gate fidelities, we perform numerical simulations using an initial coherent state $\ket{\nu}_{m}$ with amplitude $\nu=2$ (corresponding to a mean photon number $\bar{n}=4$). This choice places our setup in a slightly more demanding regime than recent continuous-variable (CV) experiments~\cite{He2023}, which reported cat-state generation with $\bar{n}\approx 2$; thus highlighting the capability of our architecture to sustain higher photon numbers.
\begin{figure*}[t]
    \centering
    \includegraphics[width=17cm]{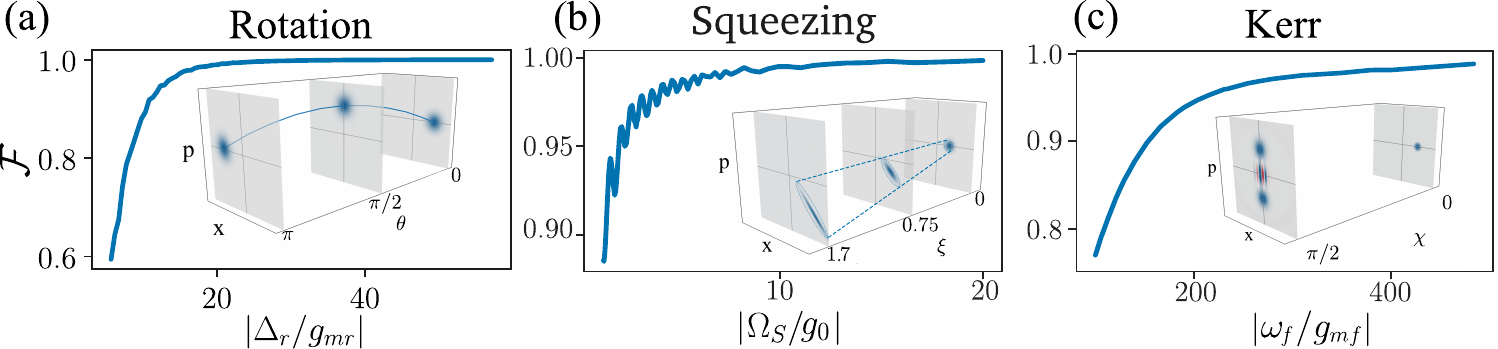}
    \caption{Numerical simulation of the fidelity of each single-mode operation as a function of the relevant device parameters. In all simulations, the system evolves under the Hamiltonian of the corresponding circuit up to the respective interaction time $\tau_{i}$ and considering the initial state $\ket{\psi_0}=\ket{\nu}_m\otimes\ket{g}_f\otimes\ket{g}_r\otimes\ket{g}_b$, with the coherent state amplitude $\nu = 2$, except for the squeezing operation in which the initial state of $F$ is $\ket{+}_f = (\ket{e}_{f} + \ket{g}_{f})/\sqrt{2}$. Then, we compare the states $\ket{\psi(\tau_i)}$ with the ideal state of each interaction. The insets show the Wigner functions for the state of $M$ during the interaction time. \textbf{(a)} For the rotation, the fidelity depends on the detuning $\Delta_r$. In the inset, we can see that the state of $M$ rotates around the phase-space origin. \textbf{(b)} For the squeezing, a behavior similar to the one in the rotation, but now, the ratio between $\Omega_S$ and $g_{mf}$ is the key to augment the fidelity. \textbf{(c)} The Kerr interaction  evolved until $\chi = \pi/2$ generates a cat state when applied in the initial coherent state and then returns to the initial state. The fidelity of the complete evolution depends on the condition $\omega_f \gg g_{mf}$.}
    \label{fig:fid_single}
\end{figure*}

\subsection{Single-mode Operations}

\textit{Rotation:} For the rotation $\mathcal{R}(\theta)$, we use a dispersive interaction between $M$ and $R$. Unlike previous approaches relying on the free evolution of the mode Hamiltonian to generate $\mathcal{R}(\theta)$~\cite{Hillmann2020}, our scheme uses $R$ to mediate the operation. This modification is essential, since allowing the mode to evolve freely under the bare Hamiltonian would result in persistent GHz oscillations, complicating the control over the rotation angle $\theta$.

Since $B$ and $F$ are not relevant for this interaction, we set them to not interact with $R$ and $M$. Therefore, the Hamiltonian for this subset of the system reads
\begin{align}\label{eq:rotation_tot}
H_{\mathcal{R}}=\omega_{m}a^{\dagger}a+\frac{\omega_{r}}{2}\sigma_{z}^{(r)}+g_{mr}\left(a^{\dagger}+a\right)\sigma_{x}^{(r)}.
\end{align}

Assuming the dispersive regime ($|\Delta_{r}|=|\omega_{m}-\omega_{r}|\gg|g_{mr}|$) and applying the rotating-wave approximation (RWA), the Hamiltonian $H_{\mathcal{R}}$ can be effectively described in the interaction picture by~\cite{James2000,Blais2004}
\begin{align}
H_{\textnormal{eff},\mathcal{R}}\approx&\frac{g_{mr}^{2}}{\Delta_{r}}(a^{\dagger}a\sigma_{gg}^{(r)}-aa^{\dagger}\sigma_{ee}^{(r)}),
\end{align}
where $\sigma_{uv}^{(r)}=\ket{u}_{r}\!\bra{v}_{r}$ ($u,v\in\{g,e\}$) denote the usual two-level operators acting on qubit $R$.

Assuming that $R$ is initially in the ground state $\ket{g}_r$, we have
\begin{align}\label{eq:rotation_eff}
H_{\textnormal{eff},\mathcal{R}}=\frac{g_{mr}^{2}}{\Delta_{r}}a^{\dagger}a, 
\end{align}
which implements the rotation $\mathcal{R}(\theta)$ with $\theta = -\tau_r g_{mr}^{2}/\Delta_{r}$, where the rotation angle can be tuned in the range $\theta \in [0, 2\pi]$ by choosing appropriate interaction time $\tau_r$ and detuning $\Delta_r$.

To evaluate the fidelity of this operation, we simulate the system evolution for $\omega_{m} =10$\,GHz, $g_{mr}=105$\,MHz, while sweeping $\omega_r$ over a range of realistic values~\cite{Kounalakis2020, Anferov2025, Roth2023, Kjaergaard2020}. For each value of $\omega_r$, the initial coherent state $\ket{\nu}_m$, with $\nu=2$ undergoes a $\pi$-rotation ($\theta = -\tau_r g_{mr}^{2}/\Delta_r = \pi$). The results are shown in Fig.~\ref{fig:fid_single}(a), where fidelity is plotted as a function of the ratio $|\Delta_r/g_{mr}|$. As expected from condition $|\Delta_r| \gg |g_{mr}|$, higher values of detuning assure better fidelity, with $|\Delta_r/g_{mr}|=57.14$ ($\omega_r=4$\,GHz) enabling $\mathcal{F}=99.98\%$ in an interaction time of $\tau_r =1.7$\,$\mu$s.

\textit{Displacement:} This operation is straightforward to implement, as it does not require the interaction between circuit elements, but only a resonant pulse applied directly to the \mbox{DC-SQUID}~\cite{Meher2022}. Therefore, we set all other components far from resonance with the mode, in such a way that the Hamiltonian reads
\begin{align}\label{eq:disp_tot}
 H_\mathcal{D} = \omega_{m} \dagg{a}a + \left( \Omega_D ae^{-\iu \omega_D t} + \Omega^*_D\dagg{a} e^{\iu \omega_D t}\right),
\end{align}
with $\Omega_D$ and $\omega_D$ the amplitude and frequency of the pulse, respectively. In the interaction picture, Eq.~(\ref{eq:disp_tot}) yields an evolution operator equal to Eq.~(\ref{sub_eq:D}) for the resonant case ($\Delta_D = \omega_{m}-\omega_D = 0$) such that $\alpha = -\iu \tau_d \Omega_D$, where $\tau_d$ is the interaction time.

Since Eq.~\eqref{eq:disp_tot} is already the exact displacement Hamiltonian, no further control techniques are required, provided our goal is to simply displace the complete state of mode $M$. In this scenario, the fidelity
of this operation, simulated for the displacement of an initial vacuum state $\ket{0}_m$ to a coherent state with $\nu = 2$, yields 100\% up to numerical errors when using resonant pulses. For these tests we used $\Omega_D = 60$~MHz for a duration $\tau_d \approx 33$~ns~\cite{Anferov2025, Sah2024}.

\textit{Squeezing:} The standard method for quadrature squeezing in superconducting circuits relies on parametric modulation or directly driving linear resonators~\cite{Zagoskin2008, Eickbusch2022}. In our case, however, the mode is not encoded in a resonator, but rather in a DC-SQUID. Since a quadratic term naturally arises from the $M$--$F$ subset, we exploit this nonlinearity to engineer the squeezing. The total Hamiltonian describing this scheme reads 
\begin{align}\label{eq:squeeze_tot}
H_\mathcal{S}&=H_{0,\mathcal{S}}+g_{mf}(a+a^{\dagger})^{2}\sigma_{x}^{(f)} \nonumber \\ &+(\Omega_{S} e^{-\iu \omega_{S}\, t}\sigma_{eg}^{(f)}+ \textnormal{H.c.}),
\end{align}
with $H_{0,\mathcal{S}} = \omega_{m}a^{\dagger}a+({\omega_{f}}/{2})\sigma_{z}^{(f)}$ the bare Hamiltonian for the mode and the fluxonium. Here $\Omega_{S}$ is the Rabi frequency of the driving field applied to the fluxonium at frequency $\omega_{S}$.

At first glance, one might attempt to obtain the squeezing in Eq.~(\ref{eq:squeeze_tot}) by imposing the condition $\omega_f = 2\omega_m$, rendering a term $\dagg{a}\dagg{a}\sigma^{(f)}_{ge}$ that is time-independent in the interaction picture. However, this condition would require a fluxonium frequency near $20$~GHz, well above the typical operation range of fluxonium qubits (0.1 -- 1~GHz)~\cite{Ding2023}, and would also conflict with the parameter regime required for the Kerr interaction discussed in the next section. Therefore, an alternative approach is necessary.

Since the coupling strength $g_{mf}$ depends explicitly on $\Phi_{\textnormal{ext}}$, we introduce a parametric modulation~\cite{Hillmann2020,Yamamoto2008, Rol2019, Ma2025} of the magnetic flux~\cite{supp_mat} yielding a time-dependent coupling $g_{mf}(t)=g_0\cos(2\omega_{1}t)$. Moving to the interaction picture with respect to $H_{0,\mathcal{S}}$, we obtain 
\begin{align}
H_{\mathcal{S},I}= & g_{mf}(t)\mathcal{Q}(t)\left(\sigma_{eg}^{(f)}e^{\iu\omega_{f}t}+\textnormal{H.c.}\right)+\Omega_{S}\sigma_{x}^{(f)},
\end{align}
where $\mathcal{Q}(t)=(a^{\dagger \;2}e^{2\iu\omega_m t}+a^2e^{-2\iu\omega_m t}+2a^{\dagger}a+\mathbb{I})$, and we use $\omega_{S}=\omega_{f}$. By adjusting $\omega_{1} = \omega_m + \omega_f/2$, the resonant quadratic terms become time-independent, whereas the off-resonant ones oscillate at frequencies of order $\omega_m,\omega_1,\omega_f$ and can be neglected via RWA provided $g_{0}\ll \omega_m,\omega_{1},\omega_{f}$,
\begin{align}
\tilde{H}_{\mathcal{S},I} & \approx\frac{g_{0}}{2}\left(aa\sigma_{ge}^{(f)}+a^{\dagger}a^{\dagger}\sigma_{eg}^{(f)}\right)+\Omega_{S}\sigma_{x}^{(f)}.
\end{align}

Finally, to separate the qubit and mode degrees of freedom, we apply an additional transformation on the qubit, $U_{1}=\exp[-i\Omega_S t\sigma_{x}^{(f)}]$, and express the Hamiltonian in the eigenbasis of $\sigma_x^{(f)}$, $\ket{\pm}_f=(\ket{e}_f\pm\ket{g}_f)/\sqrt{2}$~\cite{Prado2006}, yielding the desired effective squeezing Hamiltonian after another RWA now using $g_0 \ll \Omega_S$,
\begin{align}
    H_{\textnormal{eff},\mathcal{S}}\approx\frac{g_{0}}{2}\left(aa+a^{\dagger}a^{\dagger}\right)\left(\sigma_{++}^{(f)}-\sigma_{--}^{(f)}\right).
\end{align}

In our simulations, taking the initial coherent state $\ket{\nu}_{m}$ with $\nu=2$ for M and $\ket{+}_{f}$ for F, we obtained fidelities $\mathcal{F} = 99.8\%$ for a state squeezed by a parameter $\xi=\iu g_0 \tau_s =1.7$, corresponding to an interaction time $\tau_s = 205$~ns, with coupling strength $g_{0} = 8.3$~MHz and drive amplitude $\Omega_S = 150$~MHz, yielding a ratio $|\Omega_S/g_{0}| \approx 18$, well within current experimental capabilities~\cite{Ding2023, Felicetti2018, Bertet2005, Sah2024}. Alternatively, as shown in Fig.~\ref{fig:fid_single}(b), faster operations can be achieved by increasing $g_0$, at the cost of reduced fidelity. 

\textit{Kerr:} As mentioned earlier, the nonlinear gate required for the universal CV set could be any operation of order higher than two, such as the trisqueezing~\cite{Chang2020, Eriksson2023} or the cubic phase
gate~\cite{Gottesman2001}. Although recent works have  proposed using third-order interactions~\cite{Hillmann2022}, we focus on the fourth-order Kerr interaction~\cite{Combes2018}. This choice is advantageous because, unlike third-order interactions, the Kerr Hamiltonian conserves the number of excitations in the system. This conservation simplifies control and readout, as reading states with a high excitation number is experimentally challenging~\cite{Schuster2007}.

The Kerr interaction naturally arises in superconducting devices due to the nonlinear inductance of Josephson junctions in superconducting circuits~\cite{Kjaergaard2020}. The traditional approach to controlling its magnitude relies on circuit design, with the nonlinearities predetermined during the fabrication stage~\cite{Elliott2018, Hu2011}. However, our application requires the ability to selectively activate the Kerr interaction.

To control the Kerr effect, we again use the tunable interaction between $M$ and $F$ as in the squeezing case, but without additional driving fields or coupling moduation. The full Hamiltonian reads
\begin{align}\label{eq:Kerr_tot}
    H_{\mathcal{K}}&=H_{0,\mathcal{K}}+g_{mf}(a+a^{\dagger})^{2}\sigma_{x}^{(f)},
\end{align}
with the bare Hamiltonian $H_{0,\mathcal{K}} = \omega_{m}a^{\dagger}a+({\omega_{f}}/{2})\sigma_{z}^{(f)}$. Moving to the interaction picture with respect to $H_{0,\mathcal{K}}$, the Hamiltonian contains only oscillating terms at frequencies $\Delta_k = 2\omega_m - \omega_f$, $\Delta'_k = 2\omega_m + \omega_f$ and $\omega_f$. In the two-photon dispersive regime~\cite{Felicetti2018}, where $\omega_f, |\Delta_k|,|\Delta'_k| \gg \bar{n}|g_{mf}|$ (with $\bar{n}$ the mean number of photons in the mode), none of these terms is resonant and a second-order perturbative expansion~\cite{James2000,Felicetti2018} yields the effective Hamiltonian 
\begin{align}\label{eq:kerr_eff}
    H_{\textnormal{eff}, \mathcal{K}} = \omega' \dagg{a}a+\omega'_f\sigma_z^{(f)} + {\kappa_0} \sigma_z^{(f)} [\dagg{a}a + (\dagg{a}{a})^2],
\end{align}
where $\omega' = \omega_m +2\,g^2_{mf}(\Delta_k^{-1} -{\Delta'_k}^{-1})$ describes the frequency of the mode with an additional phase shift, which can be corrected by applying $\mathcal{R}(\theta)$ after the operation. The term $\omega'_f = \omega_f +2\,g^2_{mf}(\Delta_k^{-1} + {\Delta'_k}^{-1} + \omega_f^{-1})$ represents the shifted qubit frequency, which does not affect the Kerr interaction. Finally, the term proportional to $\kappa_0 = g^2_{mf}(\Delta_k'^{-1} + 2{\Delta^{-1}_k} + 4\omega_f^{-1})$, produces the desired Kerr interaction in $M$, with a state dependent phase shift in $F$. The evolution under this Hamiltonian for time $\tau_k$ implements the Kerr operation $\mathcal{K}(\chi)$ with $\chi = -\iu \kappa_0 \tau_k$.

Eq.~(\ref{eq:kerr_eff}) is valid only under the dispersive conditions named above, which narrows the range of parameters that can be used for the Kerr interaction. This means that, e.g., if $M$ is in the coherent state $\ket{\nu}_m$ with $\nu = 2$, the condition $\omega_f, |\Delta_k|,|\Delta'_k| \gg 4|g_{mf}|$ must be satisfied. Then, to maintain the circuit parameters within the range of $\omega_f = 400$~MHz and $\omega_m = 10$~GHz~\cite{Ding2023, Anferov2025}, it is necessary to set a small interaction strength, increasing the overall interaction time.

Considering the initial coherent state $\ket{\nu}_m$ with $\nu = 2$, the numerical simulations show that fidelities up to $98.8\%$ can be achieved by tuning the ratio  $|\omega_f/g_{mf}|$, as shown in Fig.~\ref{fig:fid_single}(c). Here, the interaction time  $\tau_k = \pi/2\kappa_0$ corresponds to the time required for the initial state to evolve into a cat state. In this setup a delicate trade-off between fidelity and gate time emerges and is mainly controlled by the $|\omega_f/g_{mf}|$ ratio. At $\omega_f/g_{mf} = 483$ (e.g.\ $g_{mf} \approx 0.83$~MHz), one obtains $\mathcal{F} = 98.8\%$ with $\tau_k \approx 228~\mu$s, while relaxing the aimed fidelity to $\mathcal{F} = 92\%$ reduces the gate time to $\tau_k \approx 27~\mu$s.

Although the Kerr gate times exceed those of the other operations in the gate set, this does not pose a fundamental limitation for the system. State-of-the-art superconducting  circuits have demonstrated coherence and relaxation times ranging from hundreds of  microseconds~\cite{VanDamme2024} to tens of milliseconds~\cite{Milul2023}, well above  both operating points identified above, confirming their experimental accessibility.

\subsection{Multi-mode Operation}
\begin{figure}
    \centering
    \includegraphics[width=\columnwidth]{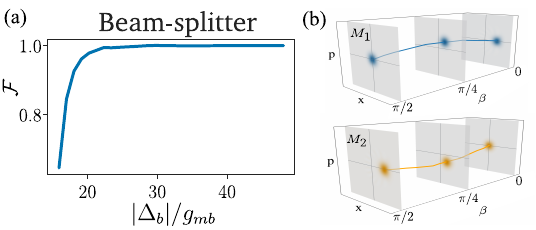}
    \caption{The beam splitter operation is achieved using a dispersive interaction between two modes mediated by the coupler $B$, see Fig. \ref{fig:system}(a). In (a) we plot the fidelity of the subspace $\ket{\psi}_{m_1}\otimes\ket{\psi}_{m_2}$ for different values of detuning between the coupler and the modes. Again, the fidelity depends on the condition $|\Delta_b| \gg g_{mb}$.  In (b), we see the Wigner plot for the states of the two modes. As can be seen, the coherent state $\ket{\psi(0)}_{m_1} = \ket{\nu}_{m_1}$, with $\nu =2$, is transferred from $M_1$ to $M_2$ after a $\pi/2$ rotation.}
    \label{fig:fid_multi}
\end{figure}

\textit{Beam splitter:}
The multimodal interaction between $M_k$ and $M_{k+1}$ will be controlled by $B_k$. This qubit will act as a coupler, blocking or allowing the exchange of information between neighboring modes~\cite{Rosario2023}. For two interacting modes, the complete Hamiltonian for the beam splitter operation is given by $H_{\mathcal{B}} = H_{0,\mathcal{B}} + V_{\mathcal{B}}$, where
\begin{align}
    H_{0,\mathcal{B}} = 
    \sum_{k=1,2}\omega_{m_k}\dagg{a}_k a_k,
    +\frac{\omega_b}{2}\sigma_z^{(b)} 
\end{align}
represents the bare Hamiltonian of the modes and the coupler, respectively, with $\omega_{m_k}$ the frequency of the $k$-th mode. The interaction between $M_1$ and $M_2$ is described by 
\begin{align}
 V_\mathcal{B} = \sum_{k=1,2} g_{m_k b}\dagg{a}_k\; \sigma_{ge}^{(b)} +\lambda \dagg{a}_1 a_2 + \textnormal{H.c.},
\end{align}
where the first term describes the capacitive coupling between $B$ and $M_k$, which we assume to be symmetric, i.e., $g_{m_1 b}=g_{m_2 b}=g_{mb}$. The second term describes the direct energy exchange between the \mbox{DC-SQUID}s due to their proximity in the circuit, with coupling strength $\lambda$ \cite{Rasmussen2021}. Usually in superconducting circuits, such direct couplings are non-tunable and  much weaker than coupler-mediated couplings ($|g_{mb}| \gg \lambda$)~\cite{Hu2023}, hence commonly referred to as parasitic couplings~\cite{Santos2023}. This non-tunability makes the native $\lambda$ term unsuitable for implementing the beam splitter operation $\mathcal{B}$. To overcome this limitation, we use an approach based on recent works~\cite{Rosario2023, Hu2023}: To control the interaction between the modes, we identify two distinct frequency values: one that blocks and another that enables information exchange between $M_1$ and $M_2$, under the assumption that the coupler remains in its ground state~\cite{Rosario2023, Hu2023}. These values are determined using the effective Hamiltonian
\begin{align}\label{eq:beamsplitter_eff}
H_{\textnormal{eff},\mathcal{B}}=\sum_{k=1,2}\frac{g_{mb}^2}{\Delta_b}\dagg{a}_k a_k + g_{\textnormal{eff},B}\left(\dagg{a}_1 a_2 +\textnormal{H.c.}\right),
\end{align}
where we assume $\omega_{m_1}=\omega_{m_2} = \omega_{m}$, call $\Delta_b=\omega_b - \omega_m$ and apply the RWA under the condition $|\Delta_b| \gg g_{mb}$. Here, the first term stands for a phase shift induced by the interaction between $M_1$, $M_2$, and $B$, which can be corrected using the previously described rotation method. The second term corresponds to the beam splitter interaction that occurs at a rate $g_{\textnormal{eff},B} =  \lambda +{g_{mb}^2/}{\Delta_b} $.

Using this effective Hamiltonian, we control the beam splitter by setting two different values of $\Delta_b$. When the beam splitter is not required, the detuning is set to $\Delta_{\textnormal{block}}=-g_{mb}^2/\lambda$, making $g_{\textnormal{eff},B} = 0$ and effectively blocking the mode interaction. When interaction is needed, $\omega_b$ can be adjusted such that $g_{\textnormal{eff},B} \neq 0$. Enabling the interaction $\mathcal{B}(\beta)$ with $\beta = \tau_b \; g_{\textnormal{eff},B}$ and $\phi = 0$.

As shown in Fig.~\ref{fig:fid_multi}, assuming the same initial coherent state and setting the interaction time to $\tau_b = \pi/2\,g_{\textnormal{eff},B}$, the beam splitter operation achieves a fidelity of $\mathcal{F}=99.91\%$ for $|\Delta_b/g_{mb}|=48$. In terms of feasible parameters for superconducting circuits~\cite{Anferov2025,Roth2023,Arute2019,Hu2023}, using $\omega_m=10$~GHz, $\omega_b = 5$~GHz, $g_{mb}=104$~MHz and $\lambda=7$~MHz, we obtain interaction times of $\tau_b=325$~ns.

\section{Conclusions}
This work presents a superconducting architecture that allows the controlled implementation of a universal gate set for continuous-variable quantum computation. By leveraging tunable interactions within a multilayer superconducting circuit, we demonstrated that all operations of the set can be
achieved with fidelities above $98\%$, within experimentally feasible parameter regimes~\cite{Anferov2025,Roth2023,Hu2023,Kounalakis2020, Kjaergaard2020, Ding2023, Felicetti2018, Bertet2005, Sah2024}. The ability to control the interaction times ensures that the interaction amplitudes can be tuned to any desired value, which is a critical feature for practical applications.

A central aspect of our proposal is the two-layer chip architecture, which is used to generate the tunable nonlinear interactions. While multilayer superconducting circuits introduce additional fabrication complexity, recent experimental demonstrations have established their viability. Multiple integration techniques including~\cite{Conner2021, Li2021, Luo2025} enable reliable inter-layer coupling with high coherence. Additionally through-silicon vias (TSVs) provide electrical access to buried layers without compromising qubit performance~\cite{Kosen2022}. Crucially, experiments show that the top layer does not significantly degrade control or readout of the bottom layer, with careful electromagnetic shielding and optimized geometries preserving individual qubit addressability and coherence times comparable to planar designs~\cite{Brecht2016, Rosenberg2017}. Given these demonstrated capabilities and the modular nature of our building block, multilayer integration represents not an obstacle but rather an opportunity—enabling compact, high-connectivity CV processors that would be geometrically hard in planar layouts.

Furthermore, our architecture overcomes key scalability limitations of previous superconducting CV implementations by employing a modular design, which allows additional modes to be incorporated without geometric constraints.

In future work, it could be important to investigate how to enhance the interaction times for the Kerr operation. Also, implementing large-scale circuits to simulate quantum CV algorithms would be an important step towards real-world applications of such circuits.

\begin{acknowledgments}
This study was financed in part, by the São Paulo Research Foundation (FAPESP), Brazil, Process Numbers 2022/00209-6, 2025/15490-0 and 2022/10218-2, by the Coordenação de Aperfeiçoamento de Pessoal de Nível Superior -- Brasil (CAPES) -- Finance Code 001, by CAPES-COFECUB (CAPES, Grant No.~88887.711967/2022-00), by the Brazilian National Council for Scientific and Technological Development -- CNPq, Grants No.~140001/2023-9, No.~405712/2023-5 and No.~311612/2021-0. A.S.M.C. acknowledges financial support from Fundação
Araucária (Project No. 305).
\end{acknowledgments}

\bibliography{refs.bib}

\end{document}

\title{\Title}

\author{Bruno A. Veloso \orcidlink{0000-0002-5935-7413}}
\email{brunoavn@df.ufscar.br}
\affiliation{\UFSCar}

\author{Ciro Micheletti~Diniz \orcidlink{0000-0002-7602-0468}}
\affiliation{\UFSCar}

\author{Luiz O. R. Solak\,\orcidlink{0000-0002-4760-3357}}
\affiliation{\UFSCar}
\affiliation{CESQ/ISIS (UMR 7006), Université de Strasbourg and CNRS, 67000 Strasbourg, France}

\author{Antonio S. M. de Castro\,\orcidlink{0000-0002-1521-9342}}
\affiliation{\UEPG}

\author{Daniel Z. Rossatto \orcidlink{0000-0001-9432-1603}}
\affiliation{\Unesp}

\author{Celso J. Villas-Bôas
\orcidlink{0000-0001-5622-786X}
}
\affiliation{\UFSCar}

\maketitle

\section{The total Hamiltonian for the M-F device}
In this section we obtain the Hamiltonian describing the M-F subset of our circuit, following the approach of Ref.~\cite{Felicetti2018} to demonstrate how a Two-Photon Rabi (TPR) Hamiltonian can be engineered. We begin by considering the Hamiltonian of the SQUID mode M,
\begin{align}
H_{m} & =E_{C}\hat{q}_{m}^{2}-2E_{J} \cos\left(\pi\frac{\Phi_{tot}}{\Phi_{0}}\right)\cos\left(\hat{\varphi}_{tot}\right).\label{eq:total}
\end{align}
Where, $E_{C}=e^{2}/C_{m}$ is the charging energy associated with
the total capacitance $C_{m}$, where $e$ is the elementary charge.
$E_{J}=I_{c}\Phi_{0}/2\pi$ is the Josephson energy of each junction,
with critical current $I_{C}$ and magnetic flux quantum $\Phi_{0}$.
$\Phi_{tot}$ denotes the total magnetic flux passing through the SQUID
loop.

The operator $\hat{q}_{m}$ represents the dimensionless charge operator
conjugate to the SQUID phase $\hat{\varphi}_{m}$ , satisfying $[\hat{\varphi}_{m},\hat{q}_{m}]=\iu$.
The total superconducting phase across the SQUID is written as $\hat{\varphi}_{tot}=\hat{\varphi}_{m}+\varphi_{DC}$
where $\varphi_{DC}$ is a classical phase bias induced by an external
current $I_{B}$ through the control line.

We now consider how the $F$ qubit modifies the mode Hamiltonian.
The total magnetic flux $\Phi_{tot}$ is not only defined by the externally applied magnetic flux $\Phi_{\textnormal{ext}}$, but also by the magnetic flux generated by $F$, denoted by $\hat{\Phi}_{F}=M_{I}I_{p}\sigma_{z}^{(f)}$, where $M_{I}$ is the mutual inductance and $I_{p}$ the persistent current amplitude in the two-level approximation. Accordingly, the total flux can be written as $ \Phi_{tot}=\Phi_{\textnormal{ext}}+\hat{\text{\ensuremath{\Phi}}}_{F}$. After transforming the flux qubit to its energy eigenbasis, $\sigma_z^{(f)}$ becomes $\sigma_x^{(f)}$. This decomposition of the total magnetic flux is central to engineering both the Kerr interaction and the squeezing Hamiltonian. We now analyze each case separately.

\subsection{Static external flux}

Let us first assume the simpler case where the external flux is static
$\Phi_{\textnormal{ext}}=\Phi_{DC}$. Assuming that the flux generated
by $F$ is small, $\langle\hat{\Phi}_{F}\rangle\ll\Phi_{0}$ and that
the SQUID operates in the transmon regime $E_{J}/E_{C}\gg1$ such
that $|\hat{\varphi}_{m}|\ll1$, we can expand the potential
of Eq.~(\ref{eq:total}) as follows

\begin{align}
&\cos\left(\frac{\pi}{\Phi_{0}}(\Phi_{DC}+\hat{\text{\ensuremath{\Phi}}}_{F})\right)  \approx\cos\left(\pi\frac{\Phi_{DC}}{\Phi_{0}}\right)-\frac{\pi}{\Phi_{0}}\sin\left(\pi\frac{\Phi_{DC}}{\Phi_{0}}\right)\hat{\text{\ensuremath{\Phi}}}_{F},\\
&\cos\left(\hat{\varphi}_{m}+\varphi_{DC}\right)  \approx\cos\left(\varphi_{DC}\right)-\sin\left(\varphi_{DC}\right)\hat{\varphi}_{m}-\frac{1}{2}\cos\left(\varphi_{DC}\right)\hat{\varphi}_{m}^{2}.\label{eq:expansion}
\end{align}

Since $\varphi_{DC}$ can be tuned via the bias current, we choose
$\varphi_{DC}=2n\pi$, with $n\in\mathbb{N}$. Under these assumptions
the Josephson potential reduces to

\begin{align}
U_{J}/E_{J} & =2\frac{\pi}{\Phi_{0}}\sin\left(\pi\frac{\Phi_{DC}}{\Phi_{0}}\right)\hat{\text{\ensuremath{\Phi}}}_{F}+\cos\left(\pi\frac{\Phi_{DC}}{\Phi_{0}}\right)\hat{\varphi}_{m}^{2}-\frac{\pi}{\Phi_{0}}\sin\left(\pi\frac{\Phi_{DC}}{\Phi_{0}}\right)\hat{\text{\ensuremath{\Phi}}}_{F}\hat{\varphi}_{m}^{2}.\label{eq:potential_dc}
\end{align}

From this potential, we can see the quadratic term $\hat{\varphi}_{m}^{2}$, which together with the $\hat{q}_{m}^{2}$ term is responsible for the harmonic behavior of our mode, along with the linear term in $\hat{\text{\ensuremath{\Phi}}}_{F}$ which produces a state-dependent energy shift proportional to $\sigma_{z}^{(f)}$. The key term, however, is the cross term$\hat{\text{\ensuremath{\Phi}}}_{F}\hat{\varphi}_{m}^{2}$,
which gives an effective two-photon interaction. 

In order to obtain the Hamiltonian in second quantization we write:
\begin{align}
H_{osc} & =E_{C}\hat{q}_{m}^{2}+E_{J}\cos\left(\pi\frac{\Phi_{DC}}{\Phi_{0}}\right)\hat{\varphi}_{m}^{2},\label{eq:total-1}
\end{align}
which corresponds to a harmonic oscillator with frequency $\omega_{m}=\sqrt{4E_{C}E_{J}\cos(\pi\Phi_{DC}/\Phi_{0})}$,
where we restrict $\cos(\pi\Phi_{DC}/\Phi_{0})>0$ to preserve the quadratic potential stability. We now can write the phase operator as

\begin{align}
\hat{\varphi}_{m} & =\left(\frac{2E_{C}}{E_{J}\cos(\pi\Phi_{DC}/\Phi_{0})}\right)^{1/4}\left(a+a^{\dagger}\right).\label{eq:potential_dc-1}
\end{align}

Applying now the same operators to the interaction term, we find the TPR Hamiltonian, 

\begin{align}
H_{m} & =\omega_{m}\left(a^{\dagger}a+\frac{1}{2}\right)+\frac{\omega_{f}}{2}\sigma_{z}^{(f)}+g_{DC}\left(a+a^{\dagger}\right)^{2}\sigma_{x}^{(f)},\label{eq:total_dc}
\end{align}
with $g_{DC}=-(\pi/4)\tan(\pi\Phi_{DC}/\Phi_{0})M_{I}I_{p}\omega_{m}/\Phi_{0}$.
Which coincides with the Hamiltonian presented in the main text.

\subsection{Dynamic external flux}

For generating the Squeezing interaction, we need to add a small time-modulation of the external magnetic flux, which reads $\Phi_{\textnormal{ext}}=\Phi_{DC}+\Phi_{AC}(t)$,
with $|\Phi_{AC}|\ll\Phi_{0}$ . This will lead to yet another expansion
from the terms $\cos\left(\pi\Phi_{\textnormal{ext}}/\Phi_{0}\right)$ and
 $\sin\left(\pi\Phi_{\textnormal{ext}}/\Phi_{0}\right).$ Expanding
and rearranging the terms lead us to the Josephson potential
\begin{align}
U_{J}/E_{J} & \approx-\left[\cos(\theta_{DC})-\frac{\pi}{\Phi_{0}}\sin(\theta_{DC})\Phi_{AC}\right]\hat{\varphi}_{m}^{2}-2\frac{\pi}{\Phi_{0}}\left[\sin(\theta_{DC})\left(1-\frac{\Phi_{AC}^{2}}{2}\right)+\frac{\pi}{\Phi_{0}}\cos(\theta_{DC})\Phi_{AC}\right]\hat{\text{\ensuremath{\Phi}}}_{F}\nonumber \\
 & +\frac{\pi}{\Phi_{0}}\left[\sin(\theta_{DC})\left(1-\frac{\Phi_{AC}^{2}}{2}\right)+\frac{\pi}{\Phi_{0}}\cos(\theta_{DC})\Phi_{AC}\right]\hat{\text{\ensuremath{\Phi}}}_{F}\hat{\varphi}_{m}^{2},
\end{align}
with $\theta_{DC}=\pi\Phi_{DC}/\Phi_{0}$. Now, we can set the constant
flux $\Phi_{DC}$in order to give us $\sin(\theta_{DC})=0$ leading
to the Hamiltonian
\begin{align}
H_{m,AC} & \approx E_{C}\hat{q}_{m}^{2}+E_{J}\hat{\varphi}_{m}^{2}+E_{J}\frac{2\pi^{2}}{\Phi_{0}^{2}}\Phi_{AC}\hat{\text{\ensuremath{\Phi}}}_{F}-E_{J}\frac{\pi^{2}}{\Phi_{0}^{2}}\Phi_{AC}\hat{\text{\ensuremath{\Phi}}}_{F}\hat{\varphi}_{m}^{2}.\label{eq:total_AC}
\end{align}

This results in a Hamiltonian in the second quantization where again we use $\hat{\varphi}_{m}=\left(2E_{C}/E_{J}\right)^{1/4}\left(a+a^{\dagger}\right)$,
to write
\begin{align}
H_{m,AC} & =\omega_{m}\left(a^{\dagger}a+\frac{1}{2}\right)+\frac{\omega_{f}}{2}\sigma_{z}^{(f)}+g_{AC}(t)\left(a+a^{\dagger}\right)^{2}\sigma_{x}^{(f)},\label{eq:total_ac_final}
\end{align}
where we define $g_{AC}(t)=-E_{J}\Phi_{AC}(t)(\pi^{2}/\Phi_{0}^2)M_{I}I_{p}\left(2E_{C}/E_{J}\right)^{1/2}$. For a sinusoidal modulation $\Phi_{AC}(t) = A\cos(2\omega_1 t)$ with amplitude $A$, we recover $g_{AC}(t) = g_0\cos(2\omega_1 t)$ 
with $g_0 = -E_J A(\pi^2/\Phi_0^2)M_I I_p(2E_C/E_J)^{1/2}$, as used in the main text. This Hamiltonian is the one we need to implement the Squeezing interaction, following the protocol described in the main text.
\bibliography{refs.bib}